\documentclass[conference]{IEEEtran}

\usepackage{graphicx}
\usepackage{amssymb}
\usepackage{amsmath}
\usepackage{caption}
\usepackage{subcaption}
\usepackage{algorithm}
\usepackage{amsthm}
\usepackage{comment}
\usepackage{cite}

\ifCLASSINFOpdf
\else
\fi
\usepackage{amsmath}

\newtheorem{theorem}{Theorem}

\begin{document}
\title{Interference Networks with Caches at Both Ends}
\providecommand{\keywords}[1]{\textbf{\textit{Index terms---}} #1}

\author{\IEEEauthorblockN{Joan S. Pujol Roig}
\IEEEauthorblockA{
Imperial College  London\\
Email: j.pujol-roig16@imperial.ac.uk}
\and
\IEEEauthorblockN{Deniz G\"{u}nd\"{u}z}
\IEEEauthorblockA{
Imperial College  London\\
d.gunduz@imperial.ac.uk}
\and
\IEEEauthorblockN{Filippo Tosato}
\IEEEauthorblockA{Toshiba Research Europe\\
filippo.tosato@toshiba-trel.com}}

\maketitle

\begin{abstract}
A $K_T \times K_R$ cache-aided wireless interference network, in which both the transmitters and the receivers are equipped with cache memories is studied. Each user requests one file from a library of $N$ popular files. The goal is to design the cache contents without the knowledge of the particular user demands, such that all possible demand combinations can be satisfied reliably over the interference channel. The achievable sum degrees-of-freedom ($\mathrm{sDoF}$) and the normalized delivery time (NDT) are studied for  centralized and decentralized network architectures, respectively. First, using a combination of interference alignment (IA), zero-forcing (ZF) and interference cancellation (IC) techniques, a novel caching and transmission scheme for centralized networks is introduced, and it is shown to improve the $\mathrm{sDoF}$ upon the state-of-the-art. Then, the NDT is studied when the content placement at the receiver caches is carried out in a decentralized manner. Our results indicate that, for this particular network architecture, caches located at the receiver side are more effective than those at the transmitter side in order to reduce the NDT.
\end{abstract}

\IEEEpeerreviewmaketitle
\section{Introduction}
It is now widely accepted that the expected wireless traffic growth \cite{cisco} can only be sustained by reducing the cell size and by enabling high-density spatial reuse of limited communication resources \cite{andrews2013seven, femtocell}. A geographical area, which is covered by one macro-cell, will require thousands of femto-cells \cite{Intel}. While this approach will reduce the congestion on the wireless access, exponentially growing traffic will increase the load on the back-haul links significantly. On the other hand, it is expected that  by 2020 video content will constitute 75$\%$ of the overall mobile data traffic \cite{cisco}. In order to exploit the features of video traffic and to overcome the bottleneck on the back-haul links, \textit{content caching} has been proposed and studied extensively in the recent years \cite{femtocaching,niesen2012caching,  maddah2015cache, niesen2014coded, maddah2015decentralized, maddah2014fundamental, ji2014average, naderializadeh2016fundamental, xu2016fundamental, cao2016fundamental, delay, gregori2016wireless, amiri2016fundamental}. The motivation for content caching is content reuse, i.e., many users request a small set of popular contents, and therefore, caching can replace, or reduce, the need for back-haul links. Cache-aided networks take advantage of distributed memories placed across the network (users devices, access points, routers,...) to bring content closer to end users. Networks have been traditionally passive in serving new user demands, that is, requests are processed only after they are received. However, in \textit{content caching}, networks try to anticipate user requests in advance and place popular contents in the available cache memories, improving the quality-of-service for all the users.\par

\begin{figure}
\centering
\includegraphics[width=0.4\textwidth]{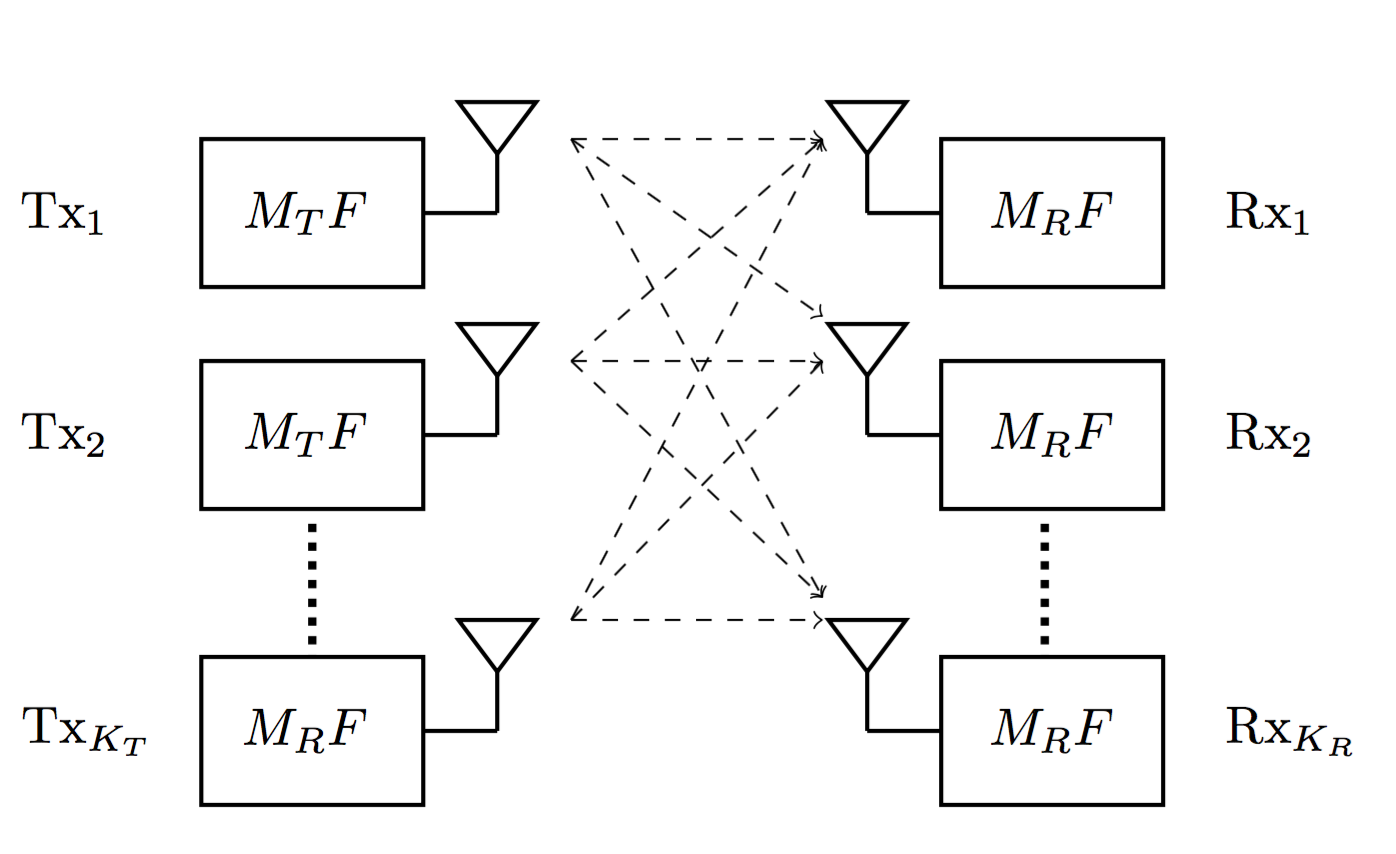}
\caption{The $K_T \times K_R$ wireless interference network with transmitter and receiver caches.}
\label{figura 1}
\end{figure}

Many models and performance measures have been considered in the literature for cache-aided network design, all showing that caching can improve the network performance drastically. In their pioneering work \cite{maddah2014fundamental}, Maddah-Ali and Niesen consider a single server, connected through an error-free shared link to $N$ users equipped with local caches. They propose a novel coded caching scheme that creates and exploits multicasting opportunities across the users, and show that it outperforms uncoded caching. In \cite{maddah2015decentralized,amiri2016decentralized} it is shown that the benefits of coded caching extends to the decentralized setting, in which users cache random bits. 
\par
The coded caching framework of \cite{maddah2014fundamental} has recently been extended to an interference network by considering cache memories at the transmitter terminals \cite{maddah2015cache}. Considering a $3\times 3$ Gaussian interference network, the authors study the achievable sum degrees-of-freedom ($\mathrm{sDoF}$). They propose a transmission scheme that exploits zero-forcing (ZF) and interference alignment (IA) techniques, leveraging content availability at the transmitter caches, in order to create different types of channels for delivering the remaining parts of the requested content. In  \cite{naderializadeh2016fundamental}, the authors extend this model to a $K_T\times K_R$  network, in which both the transmitters and receivers are equipped with cache memories. Nonetheless, in \cite{naderializadeh2016fundamental}, the authors  exploit only ZF and the receiver cache contents for interference cancellation (IC). The work in \cite{xu2016fundamental} studies the storage-latency trade-off for cache-aided wireless interference networks consisting of $K_T$ transmitters ($K_T\geq 2$) and two or three receivers. Authors introduce a new coded caching technique that leverages adjustable file splitting, and using this placement phase as well as IA and ZF techniques, they caractherize the fractional delivery time, obtaining similar results to those in \cite{maddah2015cache}, \cite{naderializadeh2016fundamental}.  A $3\times 3$ cache-aided MIMO interference network is studied in \cite{cao2016fundamental}, where the achievable fractional delivery time is characterized. Recently, the work in \cite{ursniesen} provides a constant-factor approximation of the DoF that can be obtained  in a $K_T\times K_R$  cache-aided wireless interference network with storage capabilities at both ends. The result is obtained leveraging a strategy that separates the physical and network layers. Authors show that, for this particular approach, ZF is not needed for order-optimal delivery.
Finally, in \cite{delay}, authors introduce the normalized delivery time (NDT) as a measure of performance for cache-aided networks. They characterize the NDT for $2\times 2$ and $3\times 3$ interference networks with caches only at the transmitter side. Lower-bounds on the NDT are also provided for these particular network configurations.
 \par
Similarly to \cite{naderializadeh2016fundamental} and \cite{ursniesen}, we consider a general interference network of single antenna terminals with cache capabilities at both ends. We first propose a centralized caching and delivery scheme that uses the same placement phase as in \cite{naderializadeh2016fundamental}, but leverages a combination of ZF, IA and IC. We show that the proposed scheme achieves a higher $s\mathrm{DoF}$ than the one presented in \cite{naderializadeh2016fundamental}, which does not exploit IA. Furthermore, in comparison with \cite{xu2016fundamental} and \cite{cao2016fundamental}, where authors study the case of single antenna users in a $K_T \times 3$ network and multi-antenna terminals in a $3 \times 3$ network, respectively, our proposed scheme is a general solution to the cache-aided wireless interference networks for single antenna terminals, that is, we present a valid solution for any arbitrary number of transmitters, receivers and cache sizes. Besides, in contrast to \cite{ursniesen} that presents an approximate characterization of the DoF, we provide a closed-form expression for the achievable  $s\mathrm{DoF}$.
\par
Then, we look at the decentralized caching problem. This is a more realistic scenario as we consider the transmitters as static base stations whose cache contents can be centrally coordinated, whereas the receivers correspond to mobile users whose cache contents cannot be coordinated centrally, since we do not know in advance from which base station each user will request its content during the peak traffic period. We extend the proposed delivery scheme for the centralized scenario to the decentralized setting, and characterize the corresponding achievable NDT. Previous works \cite{maddah2015decentralized, amiri2016decentralized} have focused on coding  algorithms for this type of networks in noiseless channels, but, to the best of our knowledge, this paper represents the first attempt to address the interference management problem in a decentralized cache-aided interference network.
\par

The remainder of the paper is organised as follows. The system model is introduced in Section \ref{s:System_Model}. The achievable scheme for the centralized architecture is presented in Section \ref{s:Centralized}, while it is extended to a decentralized network in Section \ref{s:Decentralized}. Finally, our conclusions are summarized in Section \ref{s:Conclusions}. 

\section{System Model}\label{s:System_Model}

We consider a wireless network  with $K_T$ transmitters $\{ \mathrm{Tx}_{1}, \mathrm{Tx}_2,\dots ,\mathrm{Tx}_{K_T}\}$ and $K_R$ receivers $\{ \mathrm{Rx}_1, \mathrm{Rx}_2,\dots ,\mathrm{Rx}_{K_R}\}$ (see Figure \ref{figura 1}), and a library of $N$ files $\{ W_{1}, W_{2},\dots ,W_{N}\}$, each of which contains $F$ bits. Each transmitter and each receiver is equipped with a cache memory of $M_TF$ and $M_RF$ bits, respectively. We impose the condition $K_TM_T+M_R\geq N$, so that the transmitters' caches are sufficient to collaboratively cache the files in the library that cannot be cached at each of the receivers. 
\par
The system operates in two phases; in the first phase, called the \textbf{\textit{placement phase}}, the transmitter and receiver caches are filled without the knowledge of the future user demands. The cache contents of $\mathrm{Tx}_i$ at the end of the placement phase is denoted by $\mathcal{P}_i$, where $|\mathcal{P}_i|\leq M_TF$, $\forall i \in [K_T] \triangleq \{1, \ldots, K_T\}$; while the cache contents of $\mathrm{Rx}_i$ is denoted by $\mathcal{Q}_i$, satisfying $|\mathcal{Q}_i|\leq M_RF$, $\forall i \in [K_R]$. In the \textbf{centralized scenario} the mapping function from the library to cache contents are known by all the nodes, while each transmitter knows only the contents of its own cache. On the other hand, in the \textbf{decentralized scenario}, coordination among users' cache contents is not possible; therefore, each user caches an equal number of bits randomly from each file in the library. \par

The receivers reveal their requests after the placement phase, and the \textbf{\textit{delivery phase}} follows. Let $W_{d_j}$ denote the file requested by $\mathrm{Rx}_j ,\ \forall j \in [K_R]$, and $\textbf{d}\triangleq [d_1,\dots, d_{K_R}] \in [N]^{K_R}$ denote the vector of all  user demands.

The delivery phase takes place over an independent and identically distributed (i.i.d.) additive white Gaussian noise interference channel. The signal received at $\mathrm{Rx}_j$ at time $t$ is:
\begin{equation}
Y_j(t)=\sum_{i=1}^{K_T}h_{ji}X_i(t)+Z_j(t),
\end{equation}
where $X_i(t)\in \mathbb{C}$ represents the signal transmitted by $\mathrm{Tx}_{i}$ at time t, $h_{ji}\in \mathbb{C}$ represents the channel coefficient between receiver $j$ and transmitter $i$, and $Z_j(t)$ is the additive Gaussian noise term at $\mathrm{Rx}_j$. We assume that the channel coefficients $\mathbf{H} \triangleq \{h_{i,j}\}_{i\in[K_R], j\in[K_T]}$, and the demand vector $\mathbf{d}$ are known by all the transmitters and receivers. 

Transmitter $\mathrm{Tx}_i$, $\forall i\in [K_T]$, maps $\mathbf{d}, \mathbf{H}$, and its own cache contents $\mathcal{P}_i$ to a channel input vector of length $L$, $\mathbf{X_i} = [X_i(1), \ldots, X_i(L)]$. We impose an average power constraint $P$ on each transmitted codeword, i.e., $\frac{1}{L}  \| \mathbf{X}_i\| ^2 \leq P$. 

Receiver $\mathrm{Rx}_i$, $\forall i\in [K_R]$, decodes its desired file $W_{d_i}$ using $\mathbf{d}, \mathbf{H}$, its own cache content $\mathcal{Q}_i$, and the corresponding channel output $\mathbf{Y}_j= [Y_i(1), \ldots, Y_j(L)]$. Let $\hat{W}_{i}$ denote its estimate of $W_{d_i}$. The sum rate of this coding scheme is $K_TF/L$, while the error probability is defined as follows:
\begin{align}
    P_e = \max_{\mathbf{d}\in [N]^{K_R}} \max_{i \in [K_R]} \mathrm{Pr}\left( \hat{W}_{i} \neq W_{d_i}\right).
\end{align}

We say that rate $R(M_T, M_R, P)$ is achievable if there exists a sequence of coding schemes with sum rate $R\left(M_T, M_R, P\right)$ that achieves a vanishing probability of error, i.e., $P_e \rightarrow 0$ as $F, L \rightarrow \infty$. We let $C(M_T, M_R,P)$ denote the maximum achievable sum rate given the cache capacities and the power constraint. Then, the sum$\mathrm{DoF}$ ($s\mathrm{DoF}$) of the system is defined as
\begin{equation}
\mathrm{sDoF}(M_T, M_R)= \liminf_{P\rightarrow \infty}\frac{C(M_T, M_R, P)}{\log(P)}.
\end{equation}
Note that, since each receiver has a single antenna, the $\mathrm{DoF}$ at each receiver is limited by one. \par

An alternative performance measure, called the normalized delivery time (NDT), is introduced in \cite{delay}, in order to account for the latency in the system. The NDT is defined as the asymptotic delivery time per bit in the high-power, long-blocklength regime, that is,
\begin{equation}
\mathrm{NDT} \triangleq \limsup_{F, P \rightarrow \infty} \frac{L}{F/\log P}.
\end{equation}
We study the achievable $\mathrm{sDoF}$ in the centralized scenario, whereas in the decentralized scenario NDT is a more appropriate performance measure since the distribution of the different pieces of a file among receiver caches is asymmetric; which means that, achievable DoF is not uniform across a file, e.g., pieces of a file stored in more than one receiver can achieve a higher DoF than those stored at only one receiver.

\section{Centralized caching}\label{s:Centralized}
In this section, we develop a coding scheme that exploits ZF, IA and IC techniques jointly in a centralized cache-aided interference network with cache memories at both ends. The same scenario is also considered in  \cite{naderializadeh2016fundamental}, where the scope is limited to ZF and IC techniques.

\begin{figure}
\centering
\includegraphics[width=0.45\textwidth]{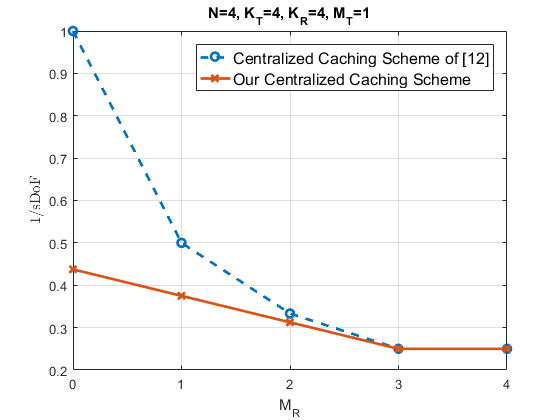}
\caption{Comparison of the inverse of the achievable $ \mathrm{sDoF}$ in the decentralized  scenario presented in this paper and the one in \cite{naderializadeh2016fundamental}.}
\label{figura 2}
\end{figure}
\subsection{Placement Phase}
We use the same algorithm as in \cite{naderializadeh2016fundamental} for the placement phase, which divides each file  into  $ \left(\begin{matrix}
K_T \\
t_T
\end{matrix} \right)\left(\begin{matrix}
K_R \\
t_R
\end{matrix} \right)$ equal-size non-overlapping subfiles, where $t_T\triangleq K_TM_T/N\in \mathbb{Z}$ and $t_R\triangleq K_TM_R/N\in \mathbb{Z}$.  For file $A \in \{W_1, \dots, W_N\}$, the subfile stored at transmitters in $X$ and receivers in $Y$ is denoted by $A_{X,Y}$.

\subsection{Delivery Algorithm}
We explain the coding scheme for the delivery phase on a simple example. Consider the following network configuration: $K_T=4,\ M_T=2,\ K_R=4,\ M_R=1,$ and $N=4$ with the file library denoted by $\{ A,B,C,D\}$. The bounds provided in \cite{naderializadeh2016fundamental} are not tight in this setting; the achievable $\mathrm{sDoF}$  is 3, while the upper-bound is 4.  For this setting we have $t_T=\frac{K_TM_T}{N}=2$ and $t_R=\frac{K_RM_R}{N}=1$. File $A$ is divided into $\left(\begin{matrix}
4 \\
2
\end{matrix}\right) \left( \begin{matrix}
4 \\
1
\end{matrix}\right) =24 $ subfiles as follows:
\begin{eqnarray*}
A_{12,1},\ A_{12,2},\ A_{12,3},\ A_{12,4},\ A_{13,1},\ A_{13,2},\ A_{13,3},\ A_{13,4},\\ A_{23,1},\ A_{23,2},\ A_{23,3},\ A_{23,4},\ A_{14,1},\ A_{14,2},\ A_{14,3},\ A_{14,4},\\ A_{24,1},\ A_{24,2},\ A_{24,3},\ A_{24,4},\ A_{34,1},\ A_{34,2},\ A_{34,3},\ A_{34,4}.
\end{eqnarray*}
The same division is applied also to files $B$, $C$ and $D$. Note that each transmitter caches half of the subfiles, whereas each receiver caches a quarter of the subfiles, of each file; resulting in all caches to be fully utilized.  \par
In the delivery phase, we assume, without loss of generality, that receivers 1, 2, 3 and 4 request files A, B, C and D, respectively (worst case demand). Each receiver already has 6 subfiles of its desired file in its cache, e.g., $\mathrm{Rx}_{1}$ has cached  subfiles $A_{12,1},\ A_{13,1},\ A_{23,1},\ A_{14,1},\ A_{24,1}$ and $A_{34,1}$. Transmitters should then provide the rest of the subfiles. 
\par
Let $A_{X,Y}^Z$ denote the subfiles of file $A$ that have been cached by transmitters in X and receivers in Y, while Z denotes the receiver at which the transmitters in X zero-force their interference when transmitting this subfile. We first focus on the transmission of the following subfiles:
\begin{equation}
\begin{split}
A_{12,2}^3,\ A_{13,2}^3,\ A_{23,2}^3,\ A_{14,2}^3,\ A_{24,2}^3, \ A_{34,2}^3,\\ 
B_{12,3}^4,\ B_{13,3}^4,\ B_{23,3}^4,\ B_{14,3}^4,\ B_{24,3}^4, \ B_{34,3}^4,\\ 
C_{12,4}^1,\ C_{13,4}^1,\ C_{23,4}^1,\ C_{14,4}^1,\ C_{24,4}^1, \ C_{34,4}^1,\\ 
D_{12,1}^2,\ D_{13,1}^2,\ D_{23,1}^2,\ D_{14,1}^3,\ D_{24,1}^2, \ D_{34,1}^2.
\end{split}
\label{eqn1}
\end{equation}

Each of these subfiles is cached by a receiver that does not request it. Naturally, each of them is going to be zero-forced at a receiver that does not have it in the cache. For example, each subfile of file $A$ in (\ref{eqn1}) is stored at $\mathrm{Rx}_{2}$ and will be zero-forced at $\mathrm{Rx}_{3}$.\par
Consider the transmission of subfile $A_{12,2}^3$. This subfile is intended to be zero-forced at $\mathrm{Rx}_{3}$; hence, $\mathrm{Tx}_{1}$ and $\mathrm{Tx}_{2}$ precode this file with the scaling factors $h_{32}$ and  $-h_{31}$, respectively. This subfile is going to be received at receivers $\mathrm{Rx}_{1}$, $\mathrm{Rx}_{2}$, $\mathrm{Rx}_{3}$ and $\mathrm{Rx}_{4}$ with the equivalent channel gains: $h_{11}h_{32} -h_{12}h_{31},\ h_{21}h_{32} -h_{22}h_{31},\ h_{31}h_{32} -h_{32}h_{31},\ h_{41}h_{32} -h_{42}h_{31}$, respectively. These channel coefficients can be restated in terms of the channel matrix minors, thus, receivers $\mathrm{Rx}_{1}$, $\mathrm{Rx}_{2}$, $\mathrm{Rx}_{3}$ and $\mathrm{Rx}_{4}$ receive  subfile $A_{12,2}^{3}$ with the scaling factors $M_{24,34},\ M_{14,34},\ 0$ and $M_{12,34}$, respectively \footnote{$M_{i,j}$ is the minor of the channel matrix $\mathbf{H}$.  The minor of a square matrix A is the determinant of the submatrix formed by deleting the i-th row and j-th column of this matrix.} .\par
\begin{figure}
\centering
\includegraphics[width=0.4\textwidth]{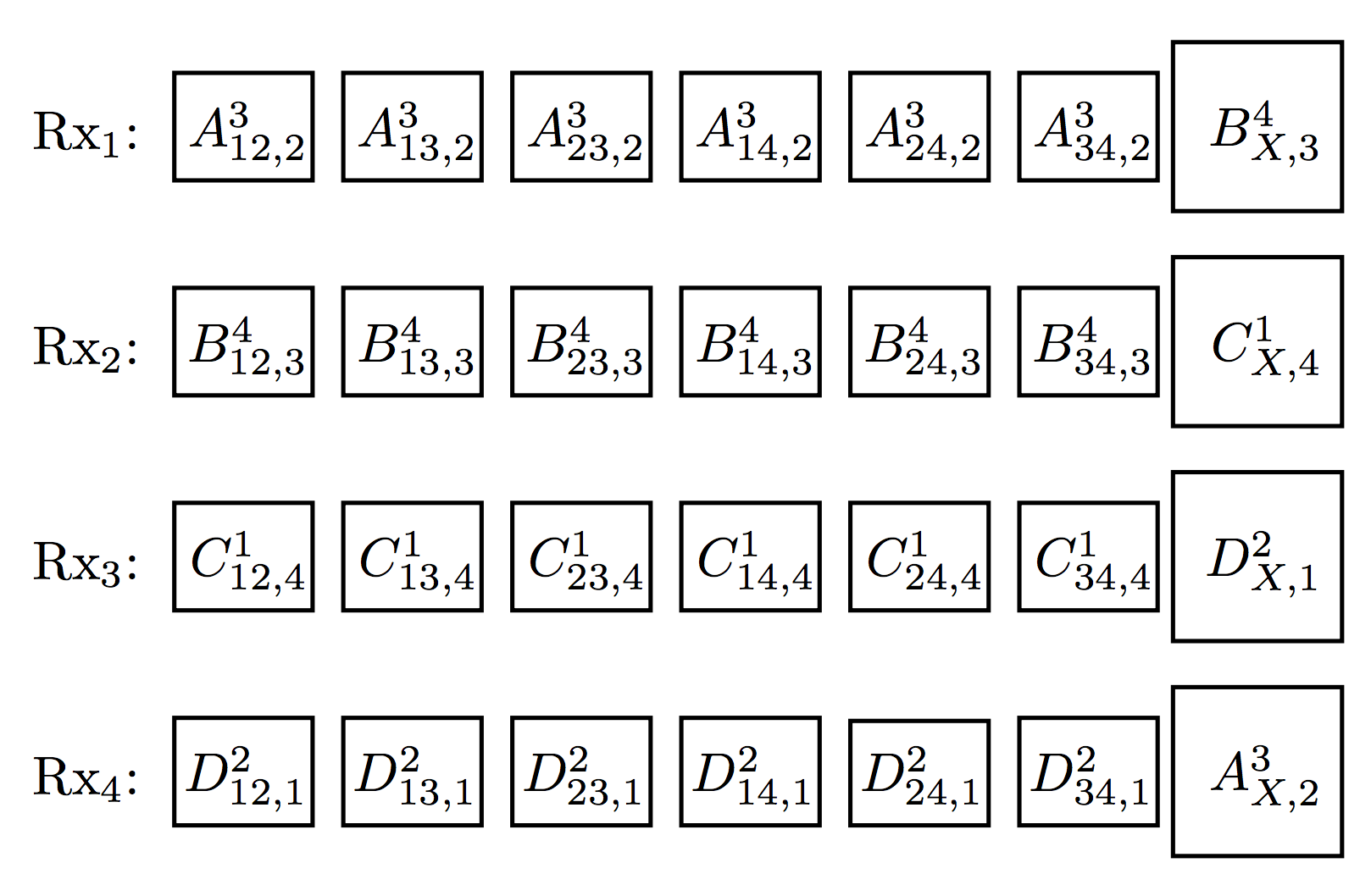}
\caption{Alignment of the subfiles in (\ref{eqn1}) at different receivers.}
\label{figura 3}
\end{figure}

We proceed by analyzing the interference caused by the transmission of subfile $A^3_{12,2}$. $\mathrm{Rx}_{1}$ requests file $A$, thus subfile $A^3_{12,2}$ is intended for this receiver. $\mathrm{Rx}_{2}$ can cancel the interference caused by this subfile, as it has already been cached by this receiver. Furthermore, this subfile is zero-forced at $\mathrm{Rx}_{3}$, meaning that no interference is caused at this particular receiver. This subfile, however, does cause interference at $\mathrm{Rx}_{4}$. We can assert that, based on the grouping of subfiles proposed above, transmission of the subfiles of file $A$ in  (\ref{eqn1}) will cause interference only at $\mathrm{Rx}_{4}$. Following a similar reasoning, the subfiles of file $B$, $C$ and $D$, respectively, will cause interference only at $\mathrm{Rx}_{1}$, $\mathrm{Rx}_{2}$ and $\mathrm{Rx}_{3}$. 
 All these subfiles are transmitted simultaneously within the same channel block, and at this point, the potential benefits of IA in this setting are conspicuous. If we accomplish aligning all the interfering signals into the same subspace at each receiver, the $\mathrm{DoF}$ will be $6/7$ per receiver (see Figure \ref{figura 3}).\par
Next, we present the construction of the IA scheme. As said before, the interference at $\mathrm{Rx}_{1}$ is caused by subfiles $B_{12,3}^4,\ B_{13,3}^4, \ B_{23,3}^4,\ B_{14,3}^4,\ B_{24,3}^4$ and $B_{34,3}^4$, which arrive at the receiver with equivalent channel coefficients $M_{23,34},\ M_{23,14},\ M_{23,23},\ M_{23,24},\ M_{23,13}$ and $M_{23,12}$, respectively. The goal is to align all these interfering signals at the receiver. To do so, \textit{real IA} techniques presented in \cite{interference} and \cite{maddah2015cache} are leveraged, which show that  transmitters are able to align the interfering subfiles at the receiver (due to space limitation, proofs are omitted and will be available in an extended version). Similarly, the interfering signals at $\mathrm{Rx}_{2}$, $\mathrm{Rx}_{3}$  and $\mathrm{Rx}_{4}$  can also be aligned. Note that  6 subfiles are transmitted and all the interfering signals collapse into one dimension (see Figure \ref{figura 3}).
\par
 By carefully grouping the remaining subfiles, a similar transmission scheme can be exploited. We now proceed with the transmission of the subfiles $A_{X,3}^4,\ B_{X,4}^1,\ C_{X,1}^2,\ D_{X,2}^3$, where $X$ represents the transmitters at which each subfile is cached. We can see that each receiver is interfered by only one of those groups of subfiles: $\mathrm{Rx}_{2}$ will  be interfered only by subfiles  $A_{X,3}^4$; subfiles in $B_{X,4}^1$ will cause interference only at $\mathrm{Rx}_{3}$, and  $C_{X,1}^2$ and $D_{X,2}^3$ will cause interference only at $\mathrm{Rx}_{4}$ and $\mathrm{Rx}_{1}$,  respectively.
Now, by leveraging the  same IA scheme explained above, we can guarantee that all the interference collide in the same sub-space.\par
Subfiles $A_{X,4}^2,B_{X,1}^3,C_{X,2}^4$ and $D_{X,3}^1$ can be transmitted over a third channel block in a similar fashion. Note that, within each channel block the transmission of the desired subfiles at each receiver takes 6/7 of the achievable unit $\mathrm{DoF}$. Hence, we can achieve a per user $\mathrm{DoF}$ of $6/7$,  while the  $s\mathrm{DoF}$ is $24/7=3.42$.\par

We highlight that the proposed mixed-approach of IA, ZF and IC implemented in this paper achieves the highests $\mathrm{sDoF}$ in the literature for the above $4\times 4$ scenario. A comparison of the achievable 1/$\mathrm{sDoF}$ for the $4\times 4$ network between our proposed scheme and the one in \cite{naderializadeh2016fundamental} for different receiver cache size $M_R$ is presented in figure \ref{figura 2} (we use the reciprocal $1/\mathrm{sDoF}$ because is a convex function of $M_T$ and $M_R$). Furthermore, our scheme does not require the use of scaling factors as in \cite{maddah2015cache}. In order to guarantee independence, different channel coefficients are exploited, thus, IA is provided by the channel for free if the proper signal constellations are chosen. It must be said that, if we had used scaling factors, a better $\mathrm{DoF}$ can potentially be achieved. For example, in the $3 \times 3$ scenario with no caches at the receivers, it is shown in \cite{maddah2015cache} that $s\mathrm{DoF} = 18/7$ is achievable, while our scheme achieves a $\mathrm{sDoF}$ of $9/4$. However, our focus here is to provide a general scheme that exploits a mixed approach of IA, ZF and IC, and the generalization of the scaling factors introduced in \cite{maddah2015cache} is cumbersome. 

\subsection{Main result}

Here we present the $\mathrm{sDoF}$ result for a general scenario. We skip the proof due to space limitations, but it follows from similar arguments as the example given above. 
\begin{theorem}
For a centralized caching network with a library of N files,  $K_T$ transmitters, and $K_R$ receivers, with cache capacities of $M_T$ and $M_R$ files, respectively, the following $\mathrm{sDoF}$ is achievable if $t_T, t_R \in \mathbb{Z}$:
\end{theorem}
\begin{equation}
\min \left\lbrace \frac{\left(\begin{matrix}
K_T\\
t_T
\end{matrix}\right)K_R}{\left(\begin{matrix}
K_T\\
t_T
\end{matrix}\right) +K_R-t_T-t_R}, K_R \right\rbrace .
\end{equation} The factor $\left(\begin{matrix}
K_T\\
t_T
\end{matrix}\right)$ represents the number of orthogonal sub-spaces that the desired signals span at each receiver ($K_R$ receivers in total). The denominator corresponds to the total number of orthogonal directions received (desired signals plus interference) $\left(\begin{matrix}
K_T\\
t_T
\end{matrix}\right)+K_R-1$, the ZF technique allows us to reduce the number of required orthogonal directions by $t_T-1$. We can further reduce it by $t_R$ thanks to the information stored in the receivers' caches.\par
Below, the implication of these results are discussed:
\begin{itemize}
\item Unlike the results presented in \cite{naderializadeh2016fundamental}, the $\mathrm{sDoF}$ achieved by our scheme is not proportional to the aggregate cache size of the network. On the contrary, caches located at the transmitters are more valuable than those at the receivers. 
\item The fact that the caches at transmitter side lead to a higher $\mathrm{sDoF}$ signifies that employing IA and ZF jointly, by exploiting transmitters' cache, is more effective than IC using receivers' cache.
\item Our results match the ones provided in \cite{naderializadeh2016fundamental} when their bound is tight ($\min\{t_T+t_R,K_R\}=K_R$). 
\end{itemize}

\section{Decentralized caching}\label{s:Decentralized}
In this section we develop a coding scheme for interference management in decentralized cache-aided wireless networks. 

\subsection{Placement Phase}\label{section1}
In the decentralized scenario, we need to leverage a placement technique that allows us to exploit broadcasting opportunities without relying on the number or the identity of the users that participate in the delivery phase. Therefore, we use a random placement technique (similarly to\cite{maddah2015decentralized} and \cite{ji2014average}), where each receiver fills its cache with randomly chosen $\frac{M_R}{N}F$ bits from each file. On the transmitter site, as before, each file is divided into $ \left(\begin{matrix}
K_T \\
t_T
\end{matrix} \right)$ disjoint partitions of equal size, each of which is stored in $t_T$ transmitters. Again, we assume $t_T,\ t_R\ \in \mathbb{Z}$. Each of these partitions can be further divided into $\sum_{j=0}^{K_R}\left(  \begin{matrix}
K_R \\
j
\end{matrix}\right)$ subfiles, each corresponding to those bits that are stored by a particular subset of the receivers. Overall, we have a total of 
\begin{equation}\label{eqn: splitting}
\left( \begin{matrix}
K_T \\
t_T
\end{matrix}\right)
 \sum_{j=0}^{K_R}\left(  \begin{matrix}
K_R \\
j
\end{matrix}\right)
\end{equation} disjoint subfiles. \par 
\begin{figure}
\centering
\includegraphics[width=0.45\textwidth]{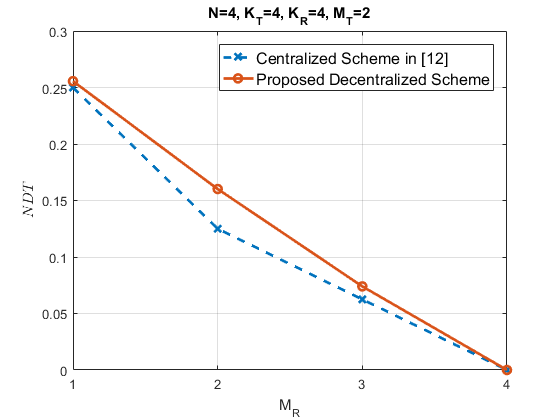}
\caption{Comparasion of the achievable NDT between the decentralized scheme presented in this paper and the centralized scheme of \cite{naderializadeh2016fundamental}.}
\label{figura 4}
\end{figure}
Unlike in the centralized scenario, subfiles  in (\ref{eqn: splitting}) are not of equal size. By the law of large numbers, the size of the subfile that has been cached by $t$ receivers can be approximated by
\begin{align}
    \left(\frac{M_R}{N}\right)^t\left( 1-\frac{M_R}{N}\right)^{K_R-t}F\ \mbox{bits}.
\end{align}
\par

\subsection{Delivery Algorithm} \label{achievable}
We consider the following setup as an example: $K_R=3,\ M_R=1,\ K_T=3,\ M_T=2$ and  $N=3$, with the library of $\{A,B,C\}$. According to the decentralized placement phase explained before, file $A$ is divided into 24 subfiles as follows:
\begin{eqnarray*}
A_{12,0},\ A_{12,1},\ A_{12,2},\ A_{12,3},\ A_{12,12},\ A_{12,13},\ A_{12,23},\ A_{12,123},\\ A_{13,0},\ A_{13,1}\ A_{13,2},\ A_{13,3}, A_{13,12},\ A_{13,13},\  A_{13,23},\ A_{13,123},\\ A_{23,0},\ A_{23,1},\ A_{23,2},\ A_{23,3},\ A_{23,12},\ A_{23,13},\ A_{23,23}, A_{23,123}
\end{eqnarray*}
The same subdivision is also applied to files $B$ and $C$. Assume that files $A$, $B$ and $C$ are requested by $\mathrm{Rx}_{1},\ \mathrm{Rx}_{2}$ and $\mathrm{Rx}_{3}$, respectively. Notice that each receiver already has 12 subfiles of each file, e.g., $\mathrm{Rx}_{1}$  has in its cache subfiles: $A_{12,1}$, $ A_{12,12}$, $A_{12,13}$, $A_{12,123}$, $A_{13,1}$, $A_{13,12}$, $A_{13,13}$, $A_{13,123}$, $A_{23,1}$, $A_{23,12}$, $A_{23,13}$ and $A_{23,123}$.\par
First, we analyze the transmission of the subfiles that have been cached by only one receiver. For these subfiles ZF, IA and IC approaches can be exploited. Consider the subfiles  $A_{12,2}$, $A_{23,2}$, $A_{13,2}$, $B_{12,3}$, $B_{23,3}$, $B_{13,3}$, $C_{12,1}$, $C_{23,1}$ and $C_{13,1}$. Each of the subfiles of file $A$ is transmitted in a different channel block (i.e., $A_{12,2}$, $B_{23,3}$ and $C_{13,1}$ in the first, $A_{23,2}$,  $B_{13,3}$ and $C_{12,1}$, in the second, and $A_{13,2}$, $B_{13,3}$ and $C_{23,1}$  in the third). Their interference is zero-forced at $\mathrm{Rx}_{3}$, by  $\mathrm{Tx}_{1}$ and $\mathrm{Tx}_{2}$. Their interference can be subtracted at $\mathrm{Rx}_{2}$, so they do not cause any interference at any receiver. Similarly, the interference from $B$'s subfiles is zero-forced at $\mathrm{Rx}_{3}$, and cancelled by $\mathrm{Rx}_{1}$. Finally, the interference from $C$'s subfiles is zero-forced at $\mathrm{Rx}_{2}$, and cancelled by $\mathrm{Rx}_{1}$. So, in this particular case, there is no need for IA  as all the signals are received  interference-free. Subfiles $A_{12,3}$,  $A_{23,3}$, $A_{13,3}$, $B_{12,3}$, $B_{23,3}$, $B_{13,3}$, $C_{12,2}$, $C_{23,2}$ and $C_{13,2}$ are transmitted in a similar fashion. The $\mathrm{sDoF}$ for these subfiles is the maximum achievable, which is 3.
\par
The transmission of subfiles $A_{12,23}$, $A_{23,23}$, $A_{13,23}$, $B_{12,13}$,  $B_{23,13}$, $B_{13,13}$, $C_{12,12}$, $C_{23,12}$ and $C_{13,12}$, cached at two receivers, is even simpler, as their interference can be cancelled by two receivers, so, they can be simply broadcasted. Interference from $A$'s subfiles can be cancelled by both $\mathrm{Rx}_{2}$ and $\mathrm{Rx}_{3}$, $B$'s subfiles by both $\mathrm{Rx}_{1}$ and $\mathrm{Rx}_{3}$, and $C$'s subfiles by both $\mathrm{Rx}_{1}$ and $\mathrm{Rx}_{2}$, so desired subfiles are received interference-free. The overall $\mathrm{sDoF}$ that can be obtained for this particular group of subfiles is 3, which, again, is the maximum.
\par
At this point, the only remaining subfiles that must be transmitted are those that have not been cached by any receiver. For transmitting these subfiles IA and ZF approaches will be simultaneously exploited. $\mathrm{Rx}_{1}$ is interested in three such subfiles, $A_{12,\emptyset}$, $A_{13,\emptyset}$ and $A_{23,\emptyset}$. At the same time, it receives interference from $B_{12,\emptyset}$, $B_{13,\emptyset}$, $B_{23,\emptyset}$, $C_{12,\emptyset}$, $C_{13,\emptyset}$ and $C_{23,\emptyset }$. However, part of these interfering signals can be eliminated using ZF as the subfiles are available at two transmitters. Following the notation of the previous section, we rewrite the signals that will be transmitted as: $ A_{12,\emptyset }^{2}$, $A_{13,\emptyset }^{2}$, $A_{23,\emptyset }^{2}$, $B_{12,\emptyset }^3$, $B_{13,\emptyset }^3$, $B_{23,\emptyset }^3$, $C_{12,\emptyset }^1$, $C_{13,\emptyset }^1$ and $C_{23,\emptyset }^1$. Thus, the only interfering signals at $\mathrm{Rx}_{1}$ are due to $B_{12,\emptyset }^3$, $B_{13,\emptyset }^3$ and $B_{23,\emptyset }^3$. We  align these interfering signals by using the same IA approach as in Section \ref{s:Centralized}. Therefore, at $\mathrm{Rx}_{1}$, subfiles of file $A$ are received in different orthogonal dimensions, while the interference signals collapse to the same orthogonal dimension; that is, each subfile carries $\frac{1}{4}$  $\mathrm{DoF}$, and the total $\mathrm{DoF}$ for $\mathrm{Rx}_{1}$ is $\frac{3}{4}$. The same methodology can be applied to the remaining subfiles cached by only a single receiver. The $\mathrm{sDoF}$ for these subfiles is found to be $\mathrm{DoF}=\frac{9}{4}$. \par
Then, the overall $\mathrm{NDT}$ can be evaluated as:
\begin{equation*}
\mathrm{NDT}=3 \left(\frac{3\cdot\left(\frac{2}{3}\right)^{3}}{\frac{9}{4}}+\frac{2 \cdot \frac{1}{3} \cdot \frac{4}{9} + \frac{1}{9} \cdot \frac{2}{3}}{3} \right) = \frac{147}{95}.
\end{equation*}
Figure \ref{figura 4} shows a comparison between the centralized delivery scheme presented in \cite{naderializadeh2016fundamental} and the proposed decentralized scheme, displaying that both schemes are very close in performance.

\subsection{Main Results}
The following theorem, stated here without proof, provides an achievable $\mathrm{NDT}$ performance for a general network.
\begin{theorem}
For a decentralized network with $K_T$ transmitters, $K_R$ receivers, a library of $N$ files, and cache size of $M_T$ and $M_R$ files at the transmitters and receivers, respectively, the following $\mathrm{NDT}$ is achievable for $t_T \in \mathbb{Z}$:
\end{theorem}
\begin{equation} \label{eq:1}
\begin{split}
K_R \sum _{t=0}^{K_R-1}\frac{\left(\left( \begin{matrix}
K_R \\
t
\end{matrix} \right)-t \right) \left(\frac{M_R}{N}\right)^t \left(1-\frac{M_R}{N}\right)^{K_R-t}}{\min \left\lbrace \frac{K_TK_R}{K_T+K_R-t_T-t},K_R\right\rbrace}\\+\frac{K_R M_T}{\min\left\lbrace K_R,\frac{K_TK_R}{K_T+K_R-t_T}\right\rbrace}\cdot \left( \max \left\lbrace M_R,1 \right\rbrace -M_R \right).
\end{split}
\end{equation}
\\
In the first part of the expression, the numerator represents the number of subfiles that need to be transmitted multiplied by the expected length. The denominator embraces the $\mathrm{sDoF}$ that can be achieved when a combination of IA, ZF, and IC are exploited. It should be noticed that the number of dimensions spanned by the interfering signals vary depending on the subfiles that are transmitted. Finally, the second part of the expression corresponds to the case with $M_R=0$. The parameter $t$ represents the gain from cancelling the interference using the receivers' cache contents, while the factor $t_T$ represents the interference cancellation due to ZF of those files that are available at more than one transmitter. 
\begin{itemize}
\item The particular value of $t_R$, as long as it is greater than $0$, does not affect the achievable $\mathrm{sDoF}$, but it reduces the size of the packets that need to be transmitted. 
\item The benefit of receiver's caches is greater than the caches at the transmitter side as it reduces the size of the subfiles that need to be transmitted, and hence, the $\mathrm{NDT}$.
\end{itemize}
The non corner points of the NDT curve, which represent the non integer values of $t_T$ and $t_R$,  can be achieved through memory-sharing \cite{maddah2014fundamental}.
\section{Conclusions}\label{s:Conclusions}
In this work, novel caching and transmission schemes for interference management in centralized and decentralized wireless cache-aided interference networks are presented. The proposed schemes make use of ZF and IA techniques as well as the users' cache contents for IC. In the case of a centralized caching network, the proposed cache-aided transmission scheme achieves the highest $\mathrm{sDoF}$ in the literature.  To the best of our knowledge, this is the first work to consider the NDT performance in a decentralized interference network, in which the cache contents of the receivers cannot be coordinated. By extending the proposed transmission strategy to the decentralized setting, we proposed an achievable NDT performance, and characterized its performance for a general network.



%
\bibliographystyle{IEEEtran}
\begin{center}
\bibliography{mybib}
\end{center}
\end{document}